\documentclass[aps,prb,superscriptaddress,twocolumn,10pt]{revtex4-1}
\usepackage[utf8]{inputenc}
\usepackage[english]{babel}
\usepackage{graphicx}
\usepackage{bm}
\usepackage{amssymb,amsmath}
\usepackage[separate-uncertainty=true]{siunitx}
\usepackage{upgreek}
\begin{document}
\newcommand{\Wcr}[0]{\ensuremath{W_{\mathrm{cr}}}}
\newcommand{\Wetu}[0]{\ensuremath{W_{\mathrm{etu}}}}
\newcommand{\mathplus}[0]{\ensuremath{+}}

\title{Analytical model for the intensity dependence of 1500~nm to 980~nm upconversion in Er$^{3+}$: a new tool for material characterization}

\author{Jeppe Christiansen}
\affiliation{Department of Physics and Astronomy, Aarhus University,\\ Ny Munkegade 120, 8000 Aarhus C, DK-Denmark}

\author{Harish Lakhotiya}
\affiliation{Department of Physics and Astronomy, Aarhus University,\\ Ny Munkegade 120, 8000 Aarhus C, DK-Denmark}
\affiliation{Interdisciplinary Nanoscience Center, Aarhus University,\\ Gustav Wieds Vej 14, DK-8000 Aarhus C, Denmark}

\author{Emil Eriksen}
\affiliation{Department of Physics and Astronomy, Aarhus University,\\ Ny Munkegade 120, 8000 Aarhus C, DK-Denmark}

\author{Søren P. Madsen}
\affiliation{Department of Engineering, Aarhus University,\\ Inge Lehmanns Gade 10, DK-8000 Aarhus C, Denmark}

\author{Peter Balling}
\affiliation{Department of Physics and Astronomy, Aarhus University,\\ Ny Munkegade 120, 8000 Aarhus C, DK-Denmark}
\affiliation{Interdisciplinary Nanoscience Center, Aarhus University,\\ Gustav Wieds Vej 14, DK-8000 Aarhus C, Denmark}

\author{Brian Julsgaard}
\email{brianj@phys.au.dk}
\affiliation{Department of Physics and Astronomy, Aarhus University,\\ Ny Munkegade 120, 8000 Aarhus C, DK-Denmark}
\affiliation{Interdisciplinary Nanoscience Center, Aarhus University,\\ Gustav Wieds Vej 14, DK-8000 Aarhus C, Denmark}

\date{\today}

\begin{abstract}
We propose a simplified rate-equation model for the \SI{1500}{\nano\meter} to \SI{980}{\nano\meter} upconversion in Er$^{3+}$. The simplifications, based on typical experimental conditions as well as on conclusions based on previously published more advanced models, enable an analytical solution of the rate equations, which reproduces known properties of upconversion. We have compared the model predictions with intensity-dependent measurements on four samples with different optical properties, such as upconversion-luminescence yield and the characteristic lifetime of the $^4I_{13/2}$ state. The saturation of the upconversion is in all cases well-described by the model over several orders of magnitude in excitation intensities. Finally, the model provides a new measure for the quality of upconverter systems based on Er$^{3+}$ -- the saturation intensity. This parameter provides valuable information on upconversion parameters such as the rates of energy-transfer upconversion and cross-relaxation. In the present investigation, we used the saturation intensity to conclude that the differences in upconversion performance of the investigated samples are mainly due to differences in the non-radiative relaxation rates.
\end{abstract}

\maketitle

\section{Introduction}
\label{Introduction}
Ions of trivalent lanthanides have shown great potential for upconversion, i.e.~the process in which two or more long-wavelength photons are combined to one of shorter wavelength \cite{Bloembergen1959, Auzel1984, Brown1969, Ovsyakin1966}. In contrast to the case of harmonic generation in nonlinear crystals, the upconversion in lanthanides is mediated by real intermediate states. This lowers the demands on the intensity and even allows upconversion of incoherent light. The optical properties of Er$^{3+}$, when embedded in crystalline or glass hosts, make it applicable in photovoltaic (PV) applications. The capacity to up-convert light with wavelengths from around \SI{1500}{\nano\meter} to \SI{980}{\nano\meter}, above the band gap of silicon, means that the energy can be absorbed and converted to electricity \cite{Trupke2006, Shalav2005, Dewild2010}. 

Upconversion is a non-linear process involving the absorption of at least two photons. Moreover, the fact that the active 4f-4f transitions are dipole forbidden necessitates high excitation intensities. However, the non-linear nature of the upconversion process facilitates an opportunity to enhance the upconversion efficiency, simply by focusing the light. In a crude model, the upconversion luminescence is proportional to the intensity of the incoming light to some power $m$, that depends on the intensity itself. In general, $m$ will have a value between one and the number of photons needed in the upconversion process \cite{Pollnau2000}. As long as $m>1$ there will be a net enhancement of the upconversion by focusing the light. Introducing metallic nanoparticles can enhance the UCL through a resonance phenomenon called a localized surface plasmon, where the incoming light interacts with the metal nanoparticle causing an increase in the near-field around the nanoparticle \cite{Schietinger2010,Han2014,Fischer2016}. In this way, we have previously obtained a seven-fold enhancement of the upconversion luminescence \cite{Harish2016}. 

To fully utilize the non-linear nature of upconversion, the saturation of the process must be understood better. Rate equations have proven to be powerful for studying lanthanide upconversion \cite{Pollnau2000,Fischer2012,MartnRodrguez2015}. The rate equations are coupled differential equations stating the probability of an erbium ion occupying one of the considered energy levels. In principle, the rate equations can easily be solved numerically. However, for a physical sound model, the parameters should be experimentally determined or at least verified, which complicates the task quite a bit. In general, rate equations for lanthanide upconversion are more complex than other typical systems, such as lasers and gain materials, due to the important processes of energy-transfer upconversion (ETU) and the reverse cross relaxation (CR), where two ions exchange energy non-radiatively \cite{Frster1948}. This further couples the rate equations, and, more importantly, introduces non-linear terms. In the literature, some authors have, nonetheless, come quite far in determining the parameters for rather complex models \cite{MartnRodrguez2015,Fischer2012}.

In the present paper, we aim to construct a simplified model for erbium-based upconversion, which provides insight into the physical processes involved. By imposing simplifications, argued for on the basis of typical experimental settings as well as parameters found in the literature \cite{Fischer2012}, an analytic expression for the UCL yield is derived, which also gives the saturation behavior. Despite the simplicity of the model, it describes the measured upconversion luminescence, including the saturation behavior, over several orders of magnitude in excitation intensity quite well for four different up-converting samples prepared in two different host materials. The model also opens new pathways for determining parameters such as the absorption cross-section and the rates of ETU and CR through relative saturation measurements instead of more involved absolute measurements.

\section{The simplified rate-equation model}
\label{sec:The_Model} 
The energy levels of lanthanide ions are in general grouped into different terms, $^{2S+1}\!L_J$, corresponding to electronic quantum numbers: $S$ for the spin, $L$ for the orbital angular momentum, and $J$ for the total angular momentum. When embedded into a host material, the crystal field causes a further Stark splitting of each term. However, this splitting is moderate in comparison with the separation between the different terms, and the relaxation among the Stark-split levels within each term is fast compared to the lifetime of the terms \cite{McCumber1964}. From a modeling perspective, it is thus a good approximation to account only for populations on a term-wise basis.
\begin{figure}
\includegraphics[width = \columnwidth]{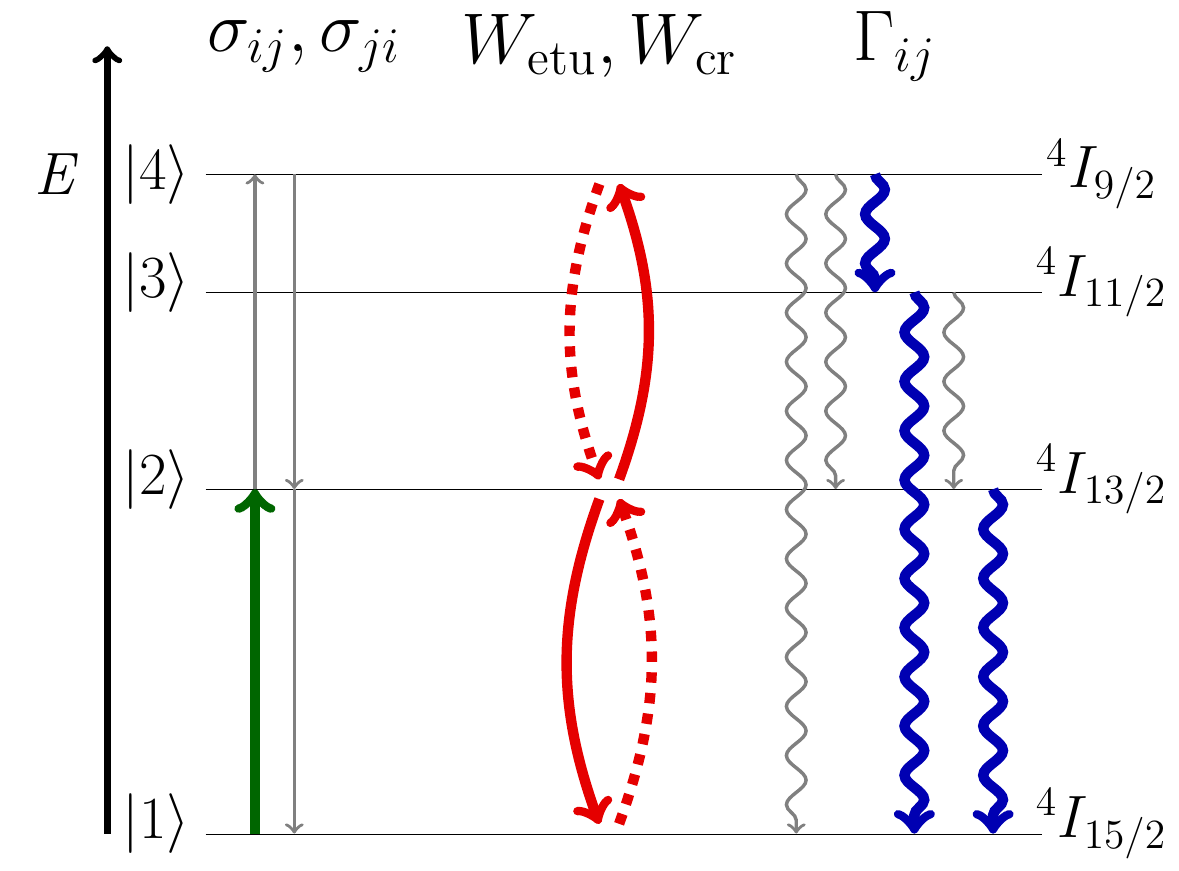} 
\caption{Illustration of the rate-equation model, with the included parameters illustrated as colored arrows, i.e absorption and stimulated emission (straight arrows), energy-transfer upconversion and cross relaxation (solid and dashed arched arrows), and radiative and non-radiative relaxation (curly arrows). The processes not included in the simplified model are depicted in gray.} 
\label{fig:REM_illustration}
\end{figure}
In the present paper, the aim is to simplify such a rate equation model as much as possible while maintaining a reasonable description of experiments. The minimal set of energy levels, which must be included in order to describe upconversion from \SI{1500}{\nano\meter} to \SI{980}{\nano\meter}, consist of the four lowest-lying terms, $^4I_{15/2}$, $^4I_{13/2}$, $^4I_{11/2}$, and $^4I_{9/2}$ (shown in Fig.~\ref{fig:REM_illustration}), since the upconversion process relies on energy transfer involving the $^4I_{9/2}$ level. In practice, the higher-lying states contribute little to the total UCL yield, and it is thus reasonable to simply truncate the rate equations to these four levels, which are henceforth denoted as levels 1, 2, 3, and 4, respectively, see Fig.~\ref{fig:REM_illustration}. Defining  $\rho_j$ as the probability that the $j$th state is populated, with $\rho_1 + \rho_2 + \rho_3 + \rho_4 = 1$ for conservation of probability, the rate equations can be written:
\begin{widetext}
\begin{equation}
\label{diff-eq:Truncated-model}
\begin{split}
  \frac{d\rho_1}{dt} = &-\frac{I}{h\nu}\sigma_{12}\rho_1 
     + [\Gamma_{21} + \frac{I}{h\nu}\sigma_{21}]\rho_2 
     + \Gamma_{31}\rho_3 + \Gamma_{41}\rho_4
     + \Wetu\rho_2^2  - \Wcr\rho_1\rho_4, \\
  \frac{d\rho_2}{dt} = &+ \frac{I}{h\nu}\sigma_{12}\rho_1 
     - [\Gamma_{21} + \frac{I}{h\nu}\{\sigma_{21} + \sigma_{24}\}]\rho_2 + 
    \Gamma_{32}\rho_3  
     + [\Gamma_{42} + \frac{I}{h\nu}\sigma_{42}]\rho_4
     - 2\Wetu\rho_2^2 + 2\Wcr\rho_1\rho_4,\\     
  \frac{d\rho_3}{dt} = &-[\Gamma_{31} + \Gamma_{32}]\rho_3
       + \Gamma_{43}\rho_4, \\
  \frac{d\rho_4}{dt} = & + \frac{I}{h\nu}\sigma_{24}\rho_2 - [\Gamma_{41} + \Gamma_{42} 
     + \Gamma_{43} + \frac{I}{h\nu}\sigma_{42}]\rho_4  + \Wetu\rho_2^2 - \Wcr\rho_1\rho_4.
  \end{split}
\end{equation}
\end{widetext}
Here $I$ is the intensity of the incoming radiation, assumed to be monochromatic at frequency $\nu$, $\sigma_{ij}$ is the cross-section for absorption or stimulated emission between levels $i$ and $j$, $\Gamma_{ij}$ is the total decay rate, radiative and non-radiative, from level $i$ to level $j$, $\Wetu$ describes the rate of ETU by F\"{o}rster resonant energy transfer with two ions in state 2 as the initial state and one ion in each of the levels 1 and 4 as the final state, and $\Wcr$ describes the reverse process cross-relaxation. Based on typical parameters for Er$^{3+}$ ions \cite{Fischer2012}, the following observations and approximations are made: (i) The excitation probability is small unless $I$ is very high. As a result, $\rho_1 \approx 1$ and $\rho_{2,3,4} \ll 1$, allowing for replacing $\rho_1$ by unity in the differential equations and for neglecting the stimulated emission term $\sigma_{21}\frac{I}{h\nu}\rho_2$, since $\sigma_{12}$ and $\sigma_{21}$ are comparable in magnitude. (ii) At the normally used concentration of Er$^{3+}$ for upconversion applications, the energy-transfer mechanism is dominating over excited-state absorption, leading to the neglect of the terms involving $\sigma_{24}$ and $\sigma_{42}$. (iii) Due to the relatively small energy difference between the states 3 and 4, the decay rate $\Gamma_{43}$, dominated by non-radiative multi-phonon relaxation, is much faster than both $\Gamma_{41}$ and $\Gamma_{42}$, which are hence neglected. (iv) $\Gamma_{31}$ exceeds $\Gamma_{32}$ by more than an order of magnitude \cite{Fischer2012}, and $\Gamma_{32}$ is neglected to further simplify the equations. With these approximations, the equations become:
\begin{equation}
\label{diff-eq:Simple-model}
  \begin{split}
  \frac{d\rho_1}{dt} = &-\sigma_{12}\frac{I}{h\nu} + \Gamma_{21}\rho_2 
     + \Gamma_{31}\rho_3  + \Wetu\rho_2^2 - \Wcr\rho_4, \\
  \frac{d\rho_2}{dt} = &+ \sigma_{12}\frac{I}{h\nu} - \Gamma_{21}\rho_2 
     - 2\Wetu\rho_2^2 + 2\Wcr\rho_4,\\     
  \frac{d\rho_3}{dt} = & -\Gamma_{31} \rho_3 + \Gamma_{43}\rho_4, \\
  \frac{d\rho_4}{dt} = & - \Gamma_{43}\rho_4 + \Wetu\rho_2^2
    - \Wcr\rho_4.
  \end{split}
\end{equation}
These rate equations, (\ref{diff-eq:Simple-model}), can be solved analytically in the steady-state regime by setting the time derivatives equal to zero. For the last equation, setting $\frac{d\rho_4}{dt} = 0 $ leads to
\begin{equation}
\label{eq:rho4_from_rho2}
\rho_4 = \frac{\Wetu \rho_2^2}{\Wcr + \Gamma_{43}}, 
\end{equation}
which can be inserted into the second equation with $\frac{d\rho_2}{dt} = 0$, leading to a quadratic equation for $\rho_2$:
\begin{equation}
  \frac{2\Wetu\Gamma_{43}}{\Wcr + \Gamma_{43}}\rho_2^2 + \Gamma_{21}\rho_2
  -\sigma_{12}\frac{I}{h\nu} = 0.
\end{equation}
The physical (positive) solution to this equation is:
\begin{equation}
\label{eq:rho2_approx_simple_model}
  \begin{split}
  \rho_2 &= \frac{\Gamma_{21}(\Wcr + \Gamma_{43})}{4\Wetu\Gamma_{43}} \left(
    \sqrt{1 + \frac{8\Wetu\Gamma_{43}\sigma_{12}I}{h\nu\Gamma_{21}^2
     (\Wcr + \Gamma_{43})}} - 1\right) \\
    &\equiv \frac{2\sigma_{12}I_{\mathrm{sat}}}{h\nu\Gamma_{21}}\left(
    \sqrt{1 + \frac{I}{I_{\mathrm{sat}}}} - 1\right),    
  \end{split}
\end{equation}
with the saturation intensity $I_{\mathrm{sat}}$ defined as
\begin{equation}
\label{eq:ISat}
  I_{\mathrm{sat}} =  \frac{h\nu\Gamma_{21}^2(\Wcr+\Gamma_{43})}
   {8\sigma_{12}\Wetu\Gamma_{43}}.
\end{equation}
It is noteworthy that this saturation intensity can be much smaller than the usual two-level-system value of $\approx \frac{\Gamma_{21}} {\sigma_{12}}$.  Setting $\frac{d\rho_3}{dt} = 0$ in the third line of Eq.~(\ref{diff-eq:Simple-model}), the population of level 3 must be
\begin{equation}
  \rho_3 = \frac{\Gamma_{43}}{\Gamma_{31}}\rho_4 = \frac{\Gamma_{43}\Wetu\rho_2^2}
   {\Gamma_{31}(\Wcr + \Gamma_{43})} 
  = \frac{h\nu\Gamma_{21}^2\rho_2^2}{8\Gamma_{31}\sigma_{12}I_{\mathrm{sat}}},
\end{equation}
where we used $\rho_4$ from Eq.~(\ref{eq:rho4_from_rho2}) in the second step. Inserting $\rho_2$ from Eq.~(\ref{eq:rho2_approx_simple_model}) leads to
\begin{equation}
\label{eq:rho3_analytical}
\rho_3 = \frac{\sigma_{12}I_{\mathrm{sat}}}{h\nu\Gamma_{31}}\left(
  1 - \sqrt{1 + \frac{I}{I_{\mathrm{sat}}}} + 
  \frac{I}{2I_{\mathrm{sat}}}\right).
\end{equation}
The rate of photon emission, $\Gamma_{\mathrm{UCL}}$, from level 3 to 1 in each ion is given by the product of the probability $\rho_3$ and the Einstein coefficient $A_{31}$ describing spontaneous emission, i.e.
\begin{equation}
\label{eq:Gamma_UCL}
\Gamma_{\mathrm{UCL}} = \Gamma_{\mathrm{eff}}  \left(
    1 - \sqrt{1 + \frac{I}{I_{\mathrm{sat}}}} + 
    \frac{I}{2I_{\mathrm{sat}}}\right),   
\end{equation}
where $\Gamma_{\mathrm{eff}}= \frac{A_{31}\sigma_{12}{I_{\mathrm{sat}}}} {h\nu\Gamma_{31}}$ was defined for brevity. In an experimental setting, the Er$^{3+}$ ions will typically be embedded in a thin film of thickness $d$ and with a concentration of $N$. If the film is much thinner than the absorption depth, and if the incoming radiation is assumed constant across a beam area $\mathcal{A}$, the total number of up-converted photons emitted per second, denoted as the UCL yield, is then given by:
\begin{equation}
\label{eq:Y_UCL_tophat}
Y_{\mathrm{UCL}} = N d \mathcal{A} \Gamma_{\mathrm{eff}}  \left\{
    1 - \sqrt{1 + \frac{I}{I_{\mathrm{sat}}}} + 
    \frac{I}{2I_{\mathrm{sat}}}\right\}.
\end{equation}
One may also consider a Gaussian intensity profile, which is relevant in typical experiments with laser excitation of the upconversion material. This leads to a distribution of intensities,
\begin{equation}
\label{eq:Gaussian}
  I(r) = \frac{2P}{\pi w_0^2} \exp\left(-\frac{2 r^2}{w_0^2}\right),
\end{equation}
where $P$ is the total beam power and $w_0$ is the beam radius where the intensity has dropped to $1/e^2$. The upconversion emission rate $\Gamma_{\mathrm{UCL}} = \Gamma_{\mathrm{UCL}}[I(r)]$ from Eq.~(\ref{eq:Gamma_UCL}) thus becomes spatially dependent through its intensity dependence, and the upconversion yield must be calculated as an integral:
\begin{equation}
\label{eq:Gaussian_integral}
  Y_{\mathrm{UCL}} = \int_0^{\infty} N d \Gamma_{\mathrm{UCL}}[I(r)] 2 \pi r dr.
\end{equation}
Since the intensity $I$ is an injective function of $r$, we can substitute $X \equiv I(r)/I_{\mathrm{sat}}$ to obtain
\begin{equation}
\label{eq:integral_I}
Y_{\mathrm{UCL}} = \frac{N d \pi w_0^2 \Gamma_{\mathrm{eff}} }{2}
\int_0^{2\bar{I}/I_{\mathrm{sat}}} \left(\frac{1-\sqrt{1+X}}{X}+\frac{1}{2}
  \right) dX,
\end{equation}
where $\bar{I} \equiv P/\pi w_0^2$ is the characteristic intensity of the Gaussian laser beam. After some algebra, one finds the result:
\begin{equation}
  \label{eq:Y_UCL_Gauss}
  \begin{split}
    Y_{\mathrm{UCL}} = Nd \pi w_0^2 \Gamma_{\mathrm{eff}} &\left\{ 1 +
       \ln
       \left[\frac{\sqrt{1+2\bar{I}/I_{\mathrm{sat}}} + 1}{2}\right]\right. \\
     &\left. + \frac{\bar{I}}{2I_{\mathrm{sat}}}
       - \sqrt{1+ 2\bar{I}/I_{\mathrm{sat}}}\right\}.
  \end{split}
\end{equation}
This result is very similar to the simple expression of Eq.~(\ref{eq:Y_UCL_tophat}) with the identification of the beam area $\mathcal{A} \rightarrow \pi w_0^2$. A comparison of the expressions is shown in Fig.~{\ref{fig:Y_UCL_illustration}(a) where $f$ is the curly brackets in either of the expressions (\ref{eq:Y_UCL_tophat}) and (\ref{eq:Y_UCL_Gauss}). The two expressions also share the asymptotic behavior
\begin{equation}
\label{eq:Y_UCL_asymptotic}
Y_{\mathrm{UCL}} = \left\{
    \begin{matrix}
      \frac{1}{8} N d \mathcal A \Gamma_{\mathrm{eff}}\left(
       \frac{I}{I_{\mathrm{sat}}}\right)^2 & \text{ when } 
      I \ll I_{\mathrm{sat}}, \\
      \frac{1}{2}N d \mathcal A \Gamma_{\mathrm{eff}}\left(
       \frac{I}{I_{\mathrm{sat}}}\right)  & \text{ when } 
      I \gg I_{\mathrm{sat}},
    \end{matrix}
    \right.
\end{equation}
provided that $\mathcal{A} = \pi w_0^2$ and $I = P/\mathcal{A}$ for the case of an incoming Gaussian beam. This reproduces the well-known quadratic behavior at low intensities, corresponding to the number of photons involved in the upconversion process, and a linear behavior in the saturated regime \cite{Pollnau2000}. We remind that the model will break down for high intensities, since $\rho_3$ was assumed much smaller than unity but it grows without limits in Eq.~(\ref{eq:rho3_analytical}) for large values of $I$. Nonetheless, with the proper attention to the range of validity, the model opens the possibility to study rate-equation parameters through Eq.~(\ref{eq:ISat}) from the saturation behavior of the UCL. This is the topic of the next sections.
\begin{figure}
\includegraphics[scale=1]{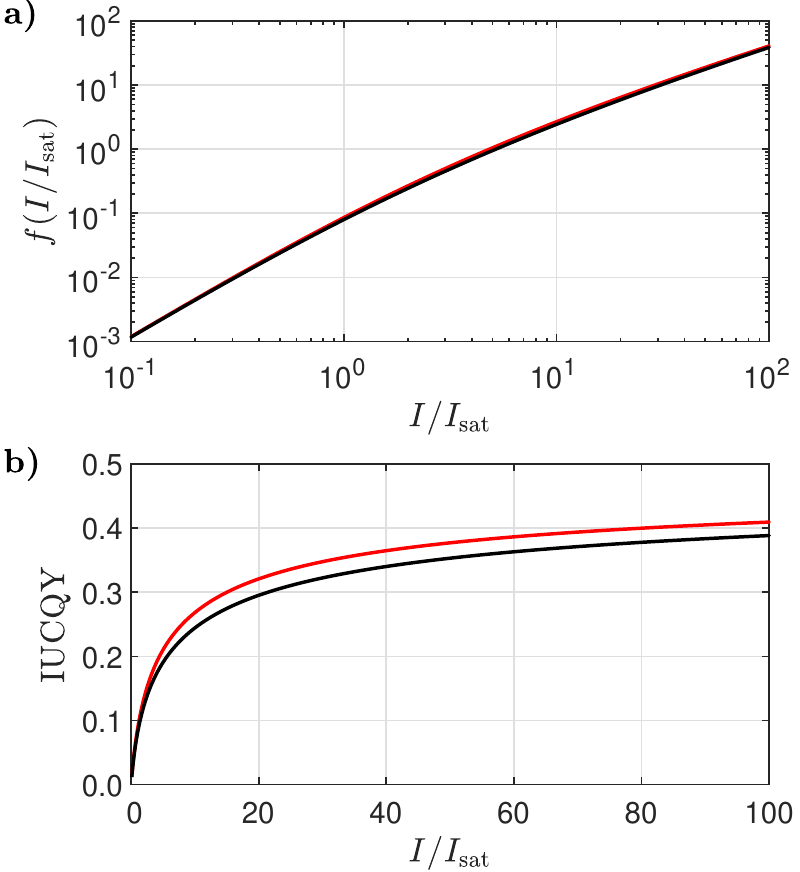}
\caption{Functional comparison of the results derived for a constant intensity profile (red) and for a Gaussian intensity profile (black). In panel a), the direct functional behavior of Eqs.~(\ref{eq:Y_UCL_tophat}) and Eq.~(\ref{eq:Y_UCL_Gauss}) is shown, whereas the IUCQY of the two expressions, Eq.~(\ref{eq:IUCQY}) assuming \SI{100}{\percent} quantum efficiency of the emitting level, is seen in panel b).}
\label{fig:Y_UCL_illustration}
\end{figure}
From the model Eqs.~(\ref{eq:Y_UCL_tophat}) and (\ref{eq:Y_UCL_Gauss}), we can estimate the internal upconversion quantum yield (IUCQY) defined as the ratio between the number of emitted and absorbed photons. In a thin slab of thickness $d$, the number of absorbed photons per second is $Y_{\mathrm{abs}} = Nd\mathcal{A}\sigma_{12}I/h\nu$, and the quantum yield then becomes
\begin{equation}
  \label{eq:IUCQY}
  \mathrm{IUCQY} = \frac{Y_{\mathrm{UCL}}}{Y_{\mathrm{abs}}}
  = \frac{A_{31}}{\Gamma_{31}} \frac{I_{\mathrm{sat}}}{I}
  f\left( \frac{I}{I_{\mathrm{sat}}}\right).
\end{equation}
The functional behavior of the IUCQY for the two intensity profile is shown in Fig.~\ref{fig:Y_UCL_illustration}(b). Due to Eq.~(\ref{eq:Y_UCL_asymptotic}), the IUCQY grows as $\mathrm{IUCQY} \approx \frac{1}{8} \frac{A_{31}}{\Gamma_{31}}\frac{I}{I_{\mathrm{sat}}}$ when $I \ll I_{\mathrm{sat}}$ and saturates to the level of $\mathrm{IUCQY} \approx \frac{1}{2}\frac{A_{31}}{\Gamma_{31}}$ when $I \gg I_{\mathrm{sat}}$. In the latter expression, the factor of $\frac{1}{2}$ reflects the fact that two photons must be absorbed in order to create one up-converted photon, and the ratio $\frac{A_{31}}{\Gamma_{31}}$ equals the quantum efficiency of radiative emission from level 3. This resembles, to a large extent, the experimentally observed behavior in Er-doped $\mathrm{NaYF_4}$ and $\mathrm{Gd_2O_2S}$ in previous investigations \cite{Fischer2014, MartnRodrguez2013}.

\section{Experimental Details}
\label{Experimental Detail}
Four samples with very different optical properties have been investigated in this work. Two of the samples consisted of chemically synthesized $\beta$-$\mathrm{NaYF_4}$ nanoparticles spin-coated in mono-layers on fused-quartz substrates. The nanoparticles were doped with Er$^{3+}$ at a concentration of approximately \SI{2.3d21}{\per\cubic\centi\meter} \cite{Harish20181}. Nanoparticles, without a shell layer and with a shell layer of \SI{10}{\nano\meter} $\mathrm{NaLuF_4}$, have been synthesized. Shell layers for rare-earth-ion doped $\mathrm{NaYF_4}$ nanoparticles are known to reduce non-radiative relaxation mediated by the surface of the nanoparticle, thus improving the upconversion efficiency \cite{Wang2010,Rabouw2018}. The synthesis and characterization of the nanoparticles are described in Ref.~\cite{Harish20181}. The remaining two samples consisted of magnetron-sputtered $\mathrm{TiO_2}$ doped with Er$^{3+}$, to a concentration of approximately \SI{4.9d21}{\per\cubic\centi\meter}, on a fused quartz substrate. The samples were sputtered at different deposition temperatures of \SI{250}{\celsius} and \SI{350}{\celsius}. The difference in temperature is known to change the upconversion properties. An increase in the density of oxygen vacancies for decreasing deposition temperature is believed to cause non-radiative relaxation, and hence quench the upconversion \cite{Harish20182}. The samples will henceforth be denoted as $\mathrm{NaYF_4}$:\SI{10}{\nano\meter}, $\mathrm{NaYF_4}$:\SI{0}{\nano\meter}, $\mathrm{TiO_2}$:D350, and $\mathrm{TiO_2}$:D250.

The samples have been investigated by steady-state UCL spectroscopy and by time-resolved photoluminescence spectroscopy. The intensity dependent UCL measurements were carried out by exciting the samples at \SI{1500}{\nano\meter} by a continuous-wave (CW) diode laser resonant with the the $^4 I_{15/2} \rightarrow ^4 I_{13/2}$ transition. The UCL was detected by a Princeton Instruments spectrograph, consisting of an Acton SP2358 monochromator and a PIXIS:100BR CCD camera. The spectra were calibrated for the spectral response by a Princeton Instrument calibration light source. The laser power was varied by inserting neutral density filters in the beam path. The laser beam-area was estimated to \SI{6e-05}{\centi\meter\squared} using the razor-blade method. The time-resolved measurements were carried out using a \SI{35}{\femto\second} pulsed Ti:sapphire laser from Spectra-Physics. The pulsed laser had a peak wavelength of \SI{800}{\nano\meter}, which made it possible to excite the $^4I_{9/2}$ level. The luminescence from the $^4I_{13/2} \rightarrow ^4I_{15/2}$ transition, at the detection wavelength of \SI{1535}{\nano\meter}, was subsequently captured by an integrated system consisting of a Princeton Instruments Acton SP2358 monochromator, a Hamamatsu R5509-73 photo-multiplier tube, and a Fast ComTech P7888-2 multi-scaler.
  
\section{Results}
\label{sec:Results}
An example of the UCL spectra measured with the highest intensity of \SI{3e02}{\watt\per\centi\meter\squared} is shown in Fig.~\ref{fig:UCL_Spectra}. The samples do not quite have the same optical properties, as relatively more luminescence from the higher excited states is seen for the $\mathrm{NaYF_4}$ samples. However, even at the highest investigated intensity, all samples emit at least an order of magnitude more light at \SI{980}{\nano\meter} compared to all other emission lines.
\begin{figure}
\includegraphics[scale=1]{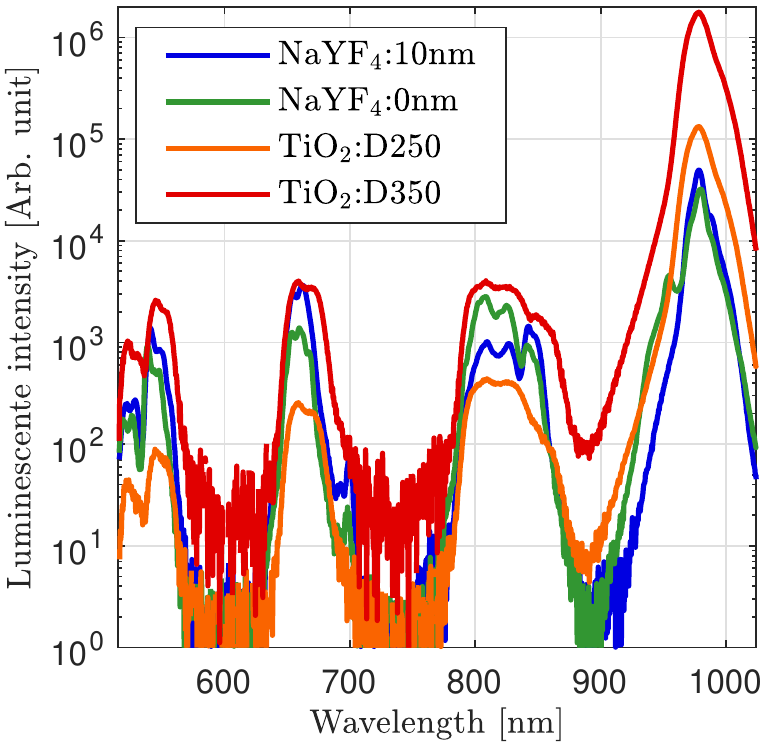} 
\caption{Example of the upconversion luminescence spectra obtained by excitation with a \SI{1500}{\nano\meter} continuous-wave laser diode. See the legend for color coding of the individual samples.}
\label{fig:UCL_Spectra}
\end{figure} 
The UCL yield was calculated by integrating the area of the UCL peak at \SI{980}{\nano\meter}, and plotted against the excitation laser intensity in Fig.~\ref{fig:Saturation_Curves} together with fits to the analytical expression in Eq.~(\ref{eq:Y_UCL_Gauss}). Notice that only two fitting parameters are needed: an amplitude/strength parameter including the pre-factor of Eq.~(\ref{eq:Y_UCL_Gauss}) as well as the experimental collection efficiency, and the saturation intensity.

The proposed model can only be expected to describe the measurements when the approximations are valid. To ensure a quantitative measure of this, data points are excluded if less than \SI{99}{\percent} of the luminescence originates from the $^4 I_{11/2}$ compared to the states, $^4S_{3/2}$ and $^4F_{9/2}$, not included in the model. None of the $\mathrm{TiO_2}$ data points have been excluded whereas some have been excluded from the $\mathrm{NaYF_4}$ samples; see the open-face data points in Fig.~\ref{fig:Saturation_Curves}(a). As seen, the model agrees well with the measurements over several orders of excitation intensity with deviations only where expected. The fitted saturation intensities are presented in Tab.~{\ref{tab:fitted_Isat}.
\begin{table}
\caption{The fitted saturation intensities in units of \si{\watt\per\centi\meter\squared} for the four investigated samples.}
\label{tab:fitted_Isat}
\begin{ruledtabular}
\begin{tabular}{l|l}
Sample & $I_{\mathrm{sat}}$ [\si{\watt\per\centi\meter\squared}] \\
\hline
$\mathrm{NaYF_4}$:\SI{10}{\nano\meter} & \num{0.5(1)} \\
$\mathrm{NaYF_4}$:\SI{0}{\nano\meter} & \num{51.3(3)} \\
$\mathrm{TiO_2}$:D350 &  \num{87.2(1)} \\
$\mathrm{TiO_2}$:D250 & \num{207.6(4)} \\
\end{tabular}
\end{ruledtabular}
\end{table}
\begin{figure}
\includegraphics[scale=1]{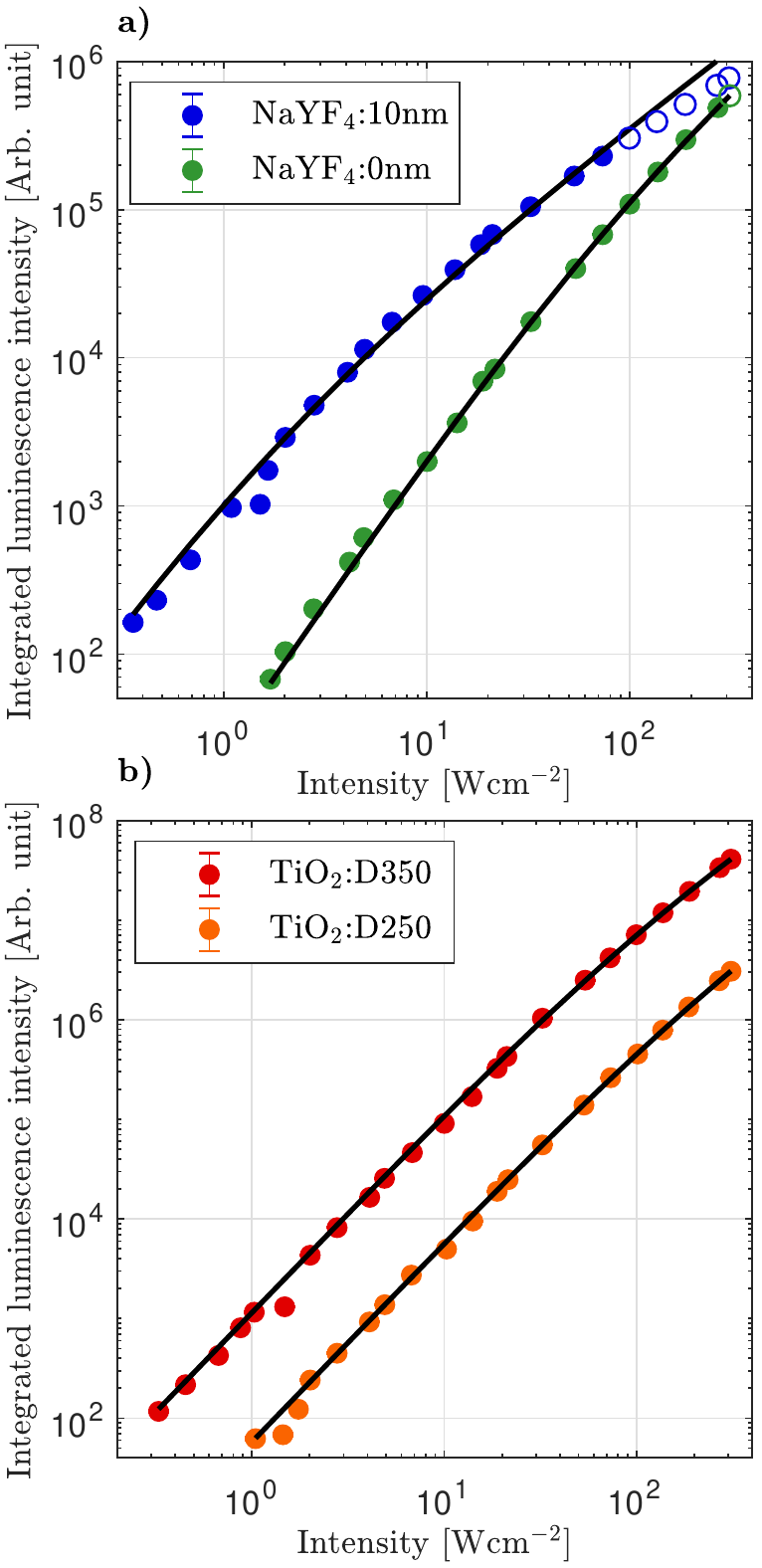} 
\caption{The measured UCL yield (colored point) and the model fit (solid black line). Open face symbols indicate that these data points have not been included in the fitting. The $\mathrm{NaYF_4}$ data are plotted in panel a) and the $\mathrm{TiO2}$ in panel b). See the legend for color coding of the individual samples.} 
\label{fig:Saturation_Curves}
\end{figure} 

The time-resolved measurements are shown in Fig.~\ref{fig:Decay_Curves}. All samples show approximately single-exponential decay with an initial rise. This can be explained by the fact that some relaxation has to occur from the excited $^4I_{9/2}$ and $^4I_{11/2}$ levels to the investigated $^4I_{13/2}$ level. The samples show very different characteristic decay times found by fitting the decay curves to functions on the form
\begin{equation}
f(t) = \left[1-a_r\exp(-g_rt)\right]\exp(-g_dt),
\label{eq:FitModel}
\end{equation}
where $a_r$ is a relative amplitude for the rise, $g_r$ is the rate of the rise, and $g_d$ is the decay rate. The time-resolved measurements were carried out at low excitation intensities to minimize the rate of energy-transfer upconversion such that $g_d$ is a good estimate of the $^4I_{13/2}$ state lifetime. Lifetimes between \SI{200}{\micro\second} and \SI{7}{\milli\second} have been determined, with the $\mathrm{NaYF_4}$:\SI{10}{\nano\meter} sample distinguishing itself with a much longer decay time than the other samples. The long decay time is probably the reason why some of the data points for this sample, fall outside the scope of the presented model and therefore have to be omitted. 
\begin{figure}
\includegraphics[scale=1]{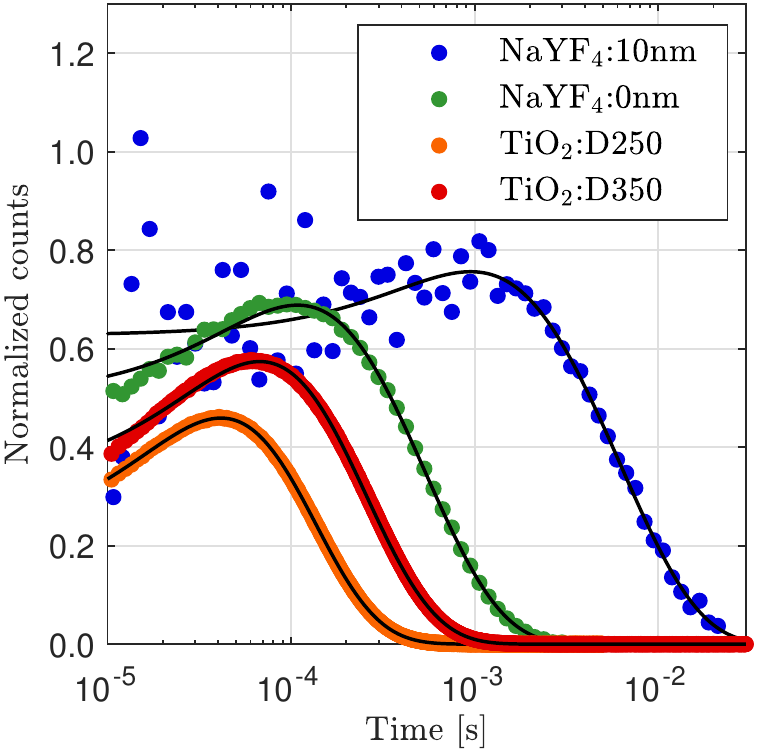} 
\caption{Measured decay curves of the $^4I_{13/2}$ state of Er$^{3+}$ for all four investigated samples (colored points) and corresponding fit (solid black line). See the legend for color coding of the individual samples.}
\label{fig:Decay_Curves}
\end{figure} 
Finally, the measured decay rate has been correlated to the fitted saturation intensity, see Fig.~\ref{fig:Isat_vs_DecayRate}. The saturation intensity is increasing strongly with an increasing decay rate in correspondence with Eq.~(\ref{eq:ISat}), assuming similar values of the unknown parameters $\Wetu$ and $\Wcr$.  
\begin{figure}
\includegraphics[scale=1]{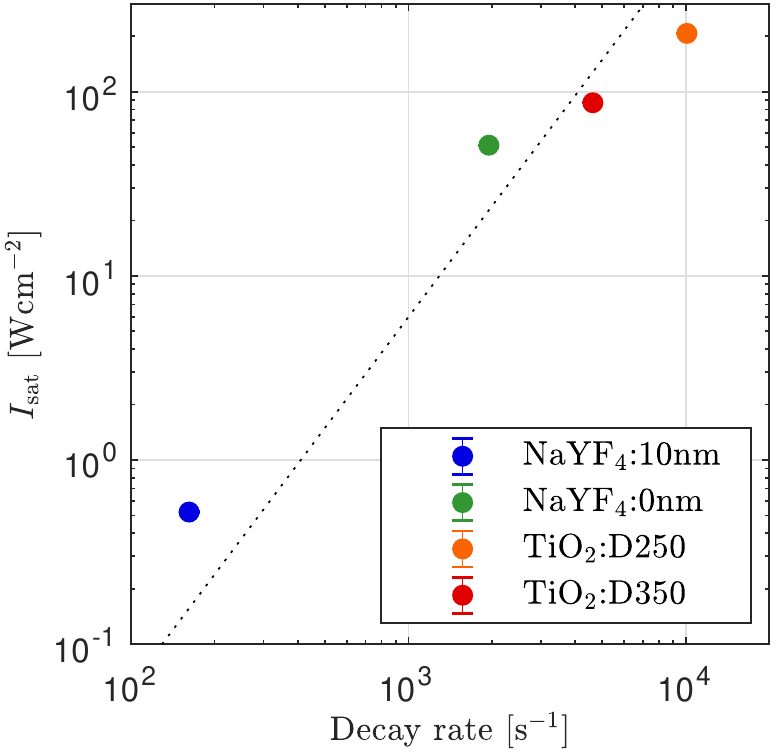}
\caption{The fitted saturation intensity plotted against the measured decay rate of the $^4I_{13/2}$ state of Er$^{3+}$. A dotted line with slope 2 has been placed to guide the eye. See the legend for color coding of the individual samples.}
\label{fig:Isat_vs_DecayRate}
\end{figure} 

\section{Discussion}
\label{sec_discussion}
The successful fitting of the saturation curves in Fig.~\ref{fig:Saturation_Curves} provides directly an experimental value for $I_{\mathrm{sat}}$, which in turn provides information about the model parameters contained in Eq.~(\ref{eq:ISat}). At first glance, this equation states that $I_{\mathrm{sat}}$ should scale quadratically with the observed decay rate. Considering Fig.~\ref{fig:Isat_vs_DecayRate}, this is almost the case, at least for the $\mathrm{NaYF_4}$ samples. However, one should bear in mind that the Er$^{3+}$ ions for the two different host materials, $\mathrm{TiO_2}$ and $\mathrm{NaYF_4}$, need not have identical absorption cross section $\sigma_{12}$, relaxation rate $\Gamma_{43}$, nor energy-transfer rates, $\Wetu$ and $\Wcr$, and these parameters may also be affected by the presence/absence of the shell for the $\mathrm{NaYF_4}$ nanoparticles or by the different deposition conditions for the $\mathrm{TiO_2}$ films. Nonetheless, the pronounced increase of $I_{\mathrm{sat}}$ with increasing decay rate still indicates that non-radiative decay channels are mainly responsible for the increased decay rate. To illustrate this point, imagine an ideal up-converting material, essentially free from non-radiative decay in order to minimize heat losses in an up-converting device. In this case, $\Gamma_{21} = A_{21}$, i.e.~the spontaneous emission rate $A_{21}$ constitutes the entire total decay rate $\Gamma_{21}$. A fast (non-radiative) decay rate $\Gamma_{43} \gg \Wcr$ is desirable though, since population would be transferred efficiently from level 4 to level 3 immediately after the energy-transfer upconversion process, thus preventing the cross-relaxation process from returning ions to level 2. In this scenario, Eq.~(\ref{eq:ISat}) reduces to $I_{\mathrm{sat}} \approx h\nu\Gamma_{21}^2/8\sigma_{12}\Wetu$. Now, suppose the material properties were varied in a way that causes changes in the dipole moment for transitions between the different energy levels, and in turn, variations in the radiative decay rate $\Gamma_{21} = A_{21}$. Such changes in the dipole coupling would also affect the absorption cross-section, $\sigma_{21} \propto A_{21}$, and the energy transfer rate, $\Wetu \propto A_{21}A_{42} \propto A_{21}^2$. The latter proportionality can be argued based on Judd-Ofeldt theory, which states that the radiative decay rates $A_{ij}$ between levels $i$ and $j$ are correlated \cite{Judd1962,Ofelt1962}. In total, for a material with radiatively limited decay rates, one would expect the scaling $I_{\mathrm{sat}} \propto \Gamma_{21}^{-1}$. Despite the material differences between the investigated samples, this is evidently not the case. The physical origin of the non-radiative decay channels will be discussed elsewhere \cite{Harish20181,Harish20182}. 

At the device level, an appropriate intensity to target is $I \approx 10 I_{\mathrm{sat}}$, see Fig.~\ref{fig:Y_UCL_illustration}(b). In comparison, on the surface of Earth, the Sun delivers $\approx$ \SI{1000}{\watt\per\meter\squared} in the entire spectrum. In a silicon solar cell, only light with a wavelength below \SI{1100}{\nano\meter} is absorbed, leaving the longer wavelengths available for upconversion processes. In a device, which combines downshifting and Er-based upconversion \cite{Goldschmidt2008,MartnRodrguez2013,MacDougall2012}, one could ideally exploit the $\approx$\SI{100}{\watt\per\meter\squared} available within the wavelength range between $\approx$ \SI{1100}{\nano\meter} and $\approx$ \SI{1550}{\nano\meter}, which in turn would require a material with $I_{\mathrm{sat}} \approx \SI{10}{\watt\per\meter\squared} = 10^{-3}\si{\watt\per\centi\meter\squared}$ if there is no concentration of the incoming radiation. It should be noted that the idea of Goldschmidt \textit{et al.}~\cite{Goldschmidt2008} includes some fluorescent concentration, and that enhancement of upconversion via plasmonic coupling to metal nanoparticles has been demonstrated \cite{Harish2016}. Still, the lowest saturation intensity of $I_{\mathrm{sat}} \approx$ \SI{0.5}{\watt\per\centi\meter\squared} in Fig.~\ref{fig:Isat_vs_DecayRate} is still more than two orders of magnitude higher than the desired intensity discussed above, which shows that also tailoring of material hosts, leading to a larger radiative decay rate, is very desirable for the exploitation of the upconversion process in solar-cell devices.

The above discussion on how physical parameters like $\Gamma_{21}$ affect $I_{\mathrm{sat}}$ can be extended to cover also the impact on the upconversion yield, $Y_{\mathrm{UCL}}$. To this end, assume that the decay rates $\Gamma_{21}$, $\Gamma_{31}$ and $A_{31}$ together with the intensity $I$ of the incoming radiation vary within some limited range, while $\Wetu$, $\Wcr$ and $\Gamma_{43}$ are assumed to vary very little. One can then see directly from Eqs.~(\ref{eq:Y_UCL_tophat}) or~(\ref{eq:Y_UCL_Gauss}) that $Y_{\mathrm{UCL}}$ is directly proportional to the quantum efficiency of $^4I_{11/2}$, defined as the ratio of radiative relaxation rate and the total relaxation rate. The dependence on $I$ and $\Gamma_{21}$ is a little more involved. The upconversion luminescence yield can be expressed commonly as:
\begin{equation}
  Y_{\mathrm{UCL}} = \frac{A_{31}}{\Gamma_{31}} \frac{Nd\mathcal{A} \sigma_{12}
     I_{\mathrm{sat}}}{h\nu} f(I/I_{sat}),
\end{equation}
where $f$ represents the curly parentheses in Eqs.~(\ref{eq:Y_UCL_tophat}) or~(\ref{eq:Y_UCL_Gauss}). For a reasonably limited intensity range, one may approximate the intensity behavior by a power law, $Y_{\mathrm{UCL}} \propto I^m$, such that the slope of the UCL versus intensity is $m$ in a double logarithmic plot. In a similar manner, the functional dependence on $\Gamma_{21}$ can be estimated by:
\begin{equation}
  \begin{split}
    \frac{d\ln Y_{\mathrm{UCL}}}{d\ln\Gamma_{21}} &= \left(
    \frac{\partial\ln Y_{\mathrm{UCL}}}{\partial\ln I_{\mathrm{sat}}}
    + \frac{\partial \ln Y_{\mathrm{UCL}}}{\partial\ln x}
    \frac{\partial \ln x}{\partial \ln I_{\mathrm{sat}}}\right)
  \frac{d \ln I_{\mathrm{sat}}}{d \ln \Gamma_{21}} \\
  &= 2(1-m).
  \end{split}
\end{equation}
For the case of Eq.~(\ref{eq:Y_UCL_tophat}) one can show that $m=1+1/\sqrt{1+I/I_{\mathrm{sat}}}$. Now, suppose the erbium-ions are driven at $m=1.5$, corrsponding to $I = 3I_{\mathrm{sat}}$, then from the above expression we obtain the estimate $Y_{\mathrm{UCL}} \propto \Gamma_{21}^{-1} = \tau_{13/2}$, where $\tau_{13/2}$ is the lifetime of level 2, denoted also by the term $^4I_{13/2}$. If, in addition, there are variations in $\Gamma_{31}$ caused solely by introduction of non-radiative decay channels (such that $A_{31}$ is constant), one would then find in total that $Y_{\mathrm{UCL}} \propto \Gamma_{31}^{-1}\Gamma_{21}^{-1} =\tau_{11/2}\tau_{13/2}$. In other words, the observed upconversion luminescence yield should scale proportionally to the product of lifetimes for the $^4I_{11/2}$ and $^4I_{13/2}$ terms. This has indeed been observed experimentally \cite{Harish20181,Harish20182}.

The above scaling laws are also very useful for simulations of upconversion luminescence yield in varying dielectric environments, since the rate equations become decoupled from the problem of calculating the local electric field by using the Maxwell equations \cite{Emil2018}.

\section{Conclusion}
\label{sec:Conclusion} 
We have presented a simplified rate-equation model, with an analytical solution, which agrees well with UCL measurements over several orders of excitation intensities. The model reproduces known features of upconversion, that is, the asymptotic behavior at low and high excitation intensities together with the transition between the two regimes due to saturation. The model provides a new way to characterize upconversion materials based on Er$^{3+}$ through the fitted saturation intensity Eq.~(\ref{eq:ISat}), which provides insight into important upconversion parameters, such as the rates of ETU and CR; parameters often difficult to measure due to the low absorption cross-section of Er$^{3+}$. Explicitly, we have used the fitted saturation intensities and the measured lifetime, of the $^4I_{13/2}$, to argue, that the differences in UCL yield observed, are mainly due to changes in the non-radiative relaxation.

\section{Acknowledgments}
This work is supported by Innovation Fund Denmark under the project "SunTune".

\medskip

\bibliography{ref}

\end{document}